\documentclass[aps,pre,12pt,onecolumn,epsf,graphicx,showpacs]{revtex4}
\usepackage{epsfig,latexsym}
\usepackage{amsmath,amscd}

\begin{document}

\title{\bf\noindent C and S induces changes in the electronic and geometric
       structure of Pd(533) and Pd(320)}

\author{Faisal Mehmood, Sergey Stolbov, and Talat S. Rahman}
\affiliation{ 116 Cardwell Hall, Department of Physics,
             Kansas State University,
             Manhattan, Kansas 66506-2600, USA}
\date{\today}

\begin{abstract}
We have performed \textit{ab initio} electronic structure
calculations of C and S adsorption on two vicinal surfaces of Pd
with different terrace geometry and width. We find both adsorbates
to induce a significant perturbation of the surface electronic and
geometric structure of Pd(533) and Pd(320). In particular C adsorbed
at the bridge site at the edge of a Pd chain in Pd(320) is found
to penetrate the surface to form a sub-surface structure. The
adsorption energies show almost linear dependence on the number of
adsorbate-metal bonds, and lie in the ranges of 5.31eV to 8.58eV for
C and 2.89eV to 5.40eV for S. A strong hybridization between
adsorbate and surface electronic states causes a large splitting of
the bands leading to a drastic decrease in the local densities of
electronic states at the Fermi-level for Pd surface atoms
neighboring the adsorbate which may poison catalytic activity of the
surface. Comparison of the results for Pd(533) with those obtained
earlier for Pd(211) suggests the local character of the impact of
the adsorbate on the geometric and electronic structures of Pd
surfaces.
\end{abstract}
\maketitle

\section{Introduction}

The elementary processes in heterogeneous catalysis, such as
adsorption of reactants, and their diffusion and reaction, are
caused by the formation, modification, or breaking of chemical bonds
between the molecules and a catalyst. Since the nature of the
chemical bonds is determined by the interplay of the electronic and
geometric structures of the catalyst surface, these characteristics
are the focus of numerous studies.

Real catalysts usually have a complex geometric structure, because
they are highly dispersed as small particles on substrates. The
surface of these particles may have microfacets with high Miller
index planes consisting of steps and kinks which may influence
significantly the reactivity of catalyst surfaces
\cite{hammer97,gambardella01,dahl99,feibelman96,hammer99,hammer01,
mavrikakis2000,loffreda03}. For instance, the N$_2$ association
reaction is extremely sensitive to the presence of steps. For
Ru(0001), its rate at the step edges is found experimentally to be
at least 9 orders of magnitude higher than that on its terrace, at
500 K \cite{dahl99}. The sticking coefficient of O$_{2}$ on the
stepped surface Ag(410) is also found to be higher than that on
Ag(100) as measured in a recent High Resolution Electron Energy Loss
(HREEL) spectroscopy experiments \cite{savio01}.

The specific role of step atoms in a chemical reaction may, however,
be more complex and may not always lead to enhanced reactivity.
Experimental observations do not show any effect of steps on the
rate of CO oxidation on Pt(335) \cite{xu93} or other metal surfaces
\cite{liu04}. Results of first principles calculation for the
reaction on Pd(211), Pd(311) \cite{hammer01} and Ir(211)
\cite{liu04} also indicate that CO oxidation barriers are
insensitive to the local surface geometry. It has also been pointed
out \cite{liu03} that dissociation reactions are always
structure-sensitive (surface steps are favored for the reactions),
while association reactions may not always be so. Furthermore, the
reactions with high valence reactant are usually more
structure-sensitive.

Apart from steps and kinks, there are other imperfections that may
affect surface reactivity. In real catalytic processes some
sub-products of reactions or other gases present in the reaction
environment may atomically adsorb on the catalyst surface and change
its reactivity. For instance, sulfur containing molecules are common
impurities in gasoline. During CO catalytic oxidation in car exhaust
refinement system, sulfur, known as inhibitor for many catalytic
reactions, adsorbs atomically on the catalyst surface and poisons
surface reactivity. In one of the earlier studies
\cite{feibelman84}, it was suggested that the depletion of the local
density of states at the Fermi-level [$N_a( E_{F})$] ({\it a}
denotes the atom contributing to LDOS) upon S adsorption can cause
poisoning of surface reactivity. Computational studies of S
adsorption on some Pd surfaces \cite{stalbov04,makkonen03,wilke96}
also show a reduction of $N_a( E_{F})$ due to a strong hybridization
of the \textit{p} states of S with the \textit{d} states of Pd.
Several experimental studies also focus on the poisoning effect of S
on metal surfaces \cite{rose01,rutkowski01,habermehl04}. Although
the rate of CO dissociation during catalytic oxidation is low, small
amount of C can atomically adsorb on the catalyst surface. Our
recent results \cite{stalbov04,makkonen03} show that C atoms
adsorbed on Pd stepped surfaces suppress substantially $N_a( E_{F})$
of the neighboring surface atoms which may be taken as indication of
poisoning. Atomic carbon adsorption on Pd surface and its poisoning
effect were also reported for the case of catalytic vinyl acetate
synthesis \cite{bowker04}. Moreover, since step sites are generally
more reactive than others on the terrace, they are also prone to
attract more impurities. The combination of steps and atomically
adsorbed impurities may thus have a significant impact on the
reactivity of catalysts.

Motivated by the above, we have investigated the effect of C and S
adsorption on the geometric and electronic structures of Pd(533), a
vicinal of Pd(111), and compared them with that on Pd(320), a
vicinal of Pd(110), for several reasons. First a comparative study
of the effect of C and S on a stepped transition metal surface would
provide a measure of the strength of the impurity substrate bond and
its impact on the surface electronic structure. Second, as the atoms
along the steps of Pd(533) and Pd(320) have local coordination of 7
and 6, respectively, a systematic study of C and S adsorption on
these two surfaces is a step towards understanding the role of
undercoordinated sites in chemical reactivity. Third, comparison of
the results on Pd(533) with those already available for Pd(211), a
smaller terrace but of similar geometry, will provide insights on
role of terrace width on the chemisorption process. Finally,
examination of the relative effect of C and S on Pd(320) will
provide the basis for comparison of results on stepped surfaces with
fcc(110) terrace geometry to those with fcc(111) terraces
\cite{stalbov04}. Since the nonequivalent atoms on a vicinal
surface, in the presence of an adsorbate, account for a complex and
inhomogeneous system, these studies are expected to have
implications also for the characteristics of nanoparticles which
contain a range of undercoordinated sites in complex environments.
Note that in an earlier study, Makkonen \textit{et al.}
\cite{makkonen03} have already presented some results for the
energetics of S adsorption on Pd(320). We have repeated these
calculations and included the results here only for completeness.

\section{Details of Computational Method}

The present first principles calculations are based of density
functional theory (DFT) \cite{kohn64} with the generalized
gradient approximation (GGA) \cite{perdew92} for the
exchange-correlation potential. Optimized surface structures and
the adsorption energies ($E_{ad}$) have been calculated using the
plane wave pseudopotential method (PWPP) \cite{payne92} with
ultrasoft pseudopotentials \cite{vanderbilt90}, while the full
potential linearized augmented plane wave (FLAPW) method
\cite{weinert82} as embodied in the {\small WIEN2K} code
\cite{blaha01} has been used to calculate the detailed electronic
structure including the local density of electronic states (LDOS)
and valence charge densities for the most interesting systems.

The fcc(533) surface consists of four-atom wide (111) terrace and a
monatomic (100) micro-facetted step edge. A perspective view of such
a surface is shown in Fig. \ref{fig533320}a. Throughout this article
we have used the following nomenclature to describe the chains of
atoms on this surface: SC (step chain) consisting of the step edge
atoms, TC1 (terrace chain 1) for the chain passing through terrace
atoms next to the step edge, TC2 (terrace chain 2) for the chain
through the terrace atoms adjacent to the corner atoms, CC (corner
chain) for the chain located between TC2 and SC, and BNN (bulk
nearest neighbor) for the ones located just underneath SC. The
Pd(533) surface was modelled by a supercell comprised of a 22 layer
slab and 12 \AA\ of vacuum.

The Pd(320) is a stepped surface with 3 atoms wide (110) terraces
and a monatomic (100)-micro-facetted step edge. Because the fcc(110)
geometry is more open than the close-packed fcc(111), a kinked
structure is formed along the step edge in Pd(320) (see Fig.
\ref{fig533320}b). For this surface, we have used notations Pd1,
Pd2, Pd3, and so on, to describe corresponding atoms in different
layers of the surface. The Pd(320) supercell included a 19 layer
slab and 11 \AA\ of vacuum. The surfaces adsorbed with S or C
contained one adsorbate atom per primitive two dimensional unit
cells, shown in Fig. \ref{fig533320}. This corresponds to the
adsorbate coverage of 1/4 monolayer (ML) for Pd(533) and 1/3 ML for
Pd(320).

For all PWPP calculations we used an energy cut-off of 290 eV, which
was found to be sufficient. A Monkhorst-Pack \textit{k}-point mesh
\cite{monkhorst76} of $(10\times 10\times 10)$, $(10\times 3\times
1)$ and $(10\times 4\times 1)$ was used to model bulk Pd, Pd(533)
and Pd(320), respectively. The bulk lattice constant was calculated
to be 3.96 \AA\ which is almost 2\% higher than the experimental
value \cite{aip70} and is typical of results obtained from DFT/GGA.
During the lattice relaxations, all atoms were allowed to fully
relax in all directions until forces on each atom were less than
0.02 eV/\AA.

For the most interesting structures, the relaxed geometries obtained
from PWPP calculations were used as input for the {\small WIEN2K}
code which further refined the geometries in a few ionic iterations.
In the FLAPW method, the LDOS and local charges are calculated
through integration over muffin-tin (MT) spheres of radius $R_{MT}$.
To analyze the effect of the adsorbate on these specific quantities
the set of $R_{MT}$ for Pd atoms should be chosen to be the same for
both the clean and the adsorbate covered surface. Ideally $R_{MT}$
should be as large as possible without causing the MT spheres to
overlap. For bulk Pd atoms, a choice of $R_{MT}=1.38$ \AA\ was found
to be optimal. However, for the Pd atoms with direct bonds to C and
S, $R_{MT}=1.08$ \AA\ (for C) and $R_{MT}=1.22$ \AA\ (for S)
provided more compatibility with the shorter C -- Pd and S -- Pd
bond lengths. For C atoms, $R_{MT}=0.926$ \AA\ and for S atoms
$R_{MT}=1.08$ \AA\ was found to be appropriate. In order to include
a reasonably large number of plane waves $(RK_{max}=7)$ with the
reduced $R_{MT}$'s for the surface atoms, we used basis sets of
1678, 3039 and 4795 LAPWs for Pd(533), S/Pd(533) and C/Pd(533),
respectively. The calculations were performed for the $(10\times
3\times 1)$ and $(6\times 3\times 1)$ \textit{k }-point mesh in the
Brillouin zone for Pd(533) and Pd(320), respectively.

\section{Results and Discussion}

\subsection{Surface Relaxations}

Optimized geometric structures of Pd(533) and Pd(320) with adsorbed
S and C were obtained for 10 possible adsorption sites on the former
and 6 sites on the latter, as shown in Fig. \ref{fig533320}. We find
the adsorbate to perturb substantially the structures of both
surfaces regardless of the site it takes. For Pd(533) the effect is
found to be most dramatic if S or C adsorb on site \# 1, as labeled
in Fig. \ref{fig533320}. Surface lattice relaxation is usually
characterized by changes in the interlayer separations ${\bf \delta
d}_{i,i+1}$, which are defined as the distances between neighboring
surface planes. In Table \ref{relax533}, we show deviations of ${\bf
\delta d}_{i,i+1}$ from those obtained for the bulk terminated
surface. In the case of clean Pd(533), these deviations characterize
the relaxation introduced upon creation of the surface from bulk
material, while for Pd(533) with adsorbed S or C, the presence of
the adsorbate also affects the nature of the surface relaxation. As
seen from Table \ref{relax533}, S and C perturb the surface
relaxation pattern of Pd(533) in very different manners. This is
caused by differences in the adsorption geometries of S and C
illustrated in Fig. \ref{figead533}. During the relaxation, the
smaller size of the C atom allows it to penetrate the surface and
form chemical bonds with the CC, SC and BNN atoms. To keep the
optimal C-Pd bond lengths, the separation between SC (layer 1) and
BNN (layer 5) atoms is expanded causing an increase in ${\bf \delta
d}_{1,2}$ and ${\bf \delta d}_{4,5}$. In contrast, the relatively
larger size of the S atoms keeps them outside the step corner. They
form bonds with SC and CC atoms and build extra S - TC2 bonds which
induce upward displacement of the TC2 atoms (layer 3) and hence an
increase in ${\bf \delta d}_{3,4}$. Interestingly, our recent
calculations performed for Pd(211) reveal a similar response of the
surface lattice to S and C adsorption \cite{stalbov04}: the
separation between SC and BNN is substantially increased upon C
adsorption and and TC is displaced upwards upon S adsorption. Note
that Pd(211) has one less chain of atoms on  its terrace as compared
to that of Pd(533). This similarity reflects the predominantly local
character of the perturbation induced by the adsorbate. The surface
atoms are displaced to form chemical bonds with the adsorbate
affecting mostly the nearest neighbors. However, some difference in
magnitudes of atomic displacements found for Pd(533) and Pd(211) may
be traced to long range interactions. It should be mentioned that
the surfaces under consideration are high Miller index surfaces with
small interlayer separations (for Pd(533) it is only 0.604 \AA).
Therefore the large percentage of interlayer separation shown in
Table \ref{relax533} corresponds to less dramatic absolute shifts.
Nevertheless, ${\bf \delta d}_{3,4} = +44.7\%$ obtained for
S/Pd(533) is 0.27 \AA\ or ~9\% of the Pd-Pd bond length which is a
significant factor and reflective of strong perturbation induced by
the adsorbate.

Lattice relaxation upon S and C adsorption has also been studied for
the kinked Pd(320) surface. The results for ${\bf \delta d}_{i,i+1}$
for S and C adsorbed at the site 1, (Fig. \ref{fig533320}) are shown
in Table \ref{relax320}. In general, the calculated relaxation
pattern $\left( --+-+\right) $ for clean Pd(320) matches well with
LEED results \cite{pussi04}. As in the case of Pd(533), the size and
chemical composition of the adsorbate atom have a remarkable effect
on the relaxation patterns for S/Pd(320) and C/Pd(320) which are
strikingly different. C atoms penetrate the kink site and form bonds
with five neighboring surface atoms including Pd1, two Pd3, Pd5 and
Pd6. Optimization of bond length causes the displacements of all
these atoms giving rise to a significant multilayer relaxation. The
S atoms, on the other hand, stay above the kink site and cause a
substantial displacement only for Pd1 and Pd5 atoms. Also for the
case of S on Pd(320), the surface relaxations reported here are in
agreement with those obtained earlier by Makkonen \textit{et al.}
\cite{makkonen03}.

An unexpected result has also been revealed for carbon adsorption at
the bridge site (C5) between Pd1 and Pd2 on Pd(320). The adsorbed C
atom is found to penetrate the surface spontaneously by pushing the
Pd1 atom away from Pd2 and diffusing between Pd1 and Pd2 followed by
backward displacement of Pd1 to restore the Pd1 - Pd2 bond. As a
result, C forms a sub-surface structure in which it has chemical
bonds with 6 neighboring Pd atoms. The initial and final
(equilibrium) geometries of this adsorption are shown in Fig.
\ref{figead320}. This also induces a large outward displacement of
Pd1 resulting in a significant (38.1\%) increase in ${\bf \delta
d}_{1,2}$ (see Table \ref{relax320}).

\subsection{Adsorption Energies}

For each adsorption site shown in Fig. \ref{fig533320}, we have
calculated the adsorption energy $E_{ad}$ for both S and C on
Pd(533) C on Pd(320) and listed them in Table \ref{ead533}. We find
that for all sites under consideration carbon has higher adsorption
energy than sulfur. This reflects the difference in the nature of
the chemical bonding for S and C to be discussed in the next
subsection. The $E_{ad}$ values spread over the range of 4.13 eV to
5.40 eV for S and 5.31 to 8.44 eV for C atoms. Similar is the range
of $E_{ad}$ for adsorption on Pd(320) (Table \ref{ead533}) for C and
Ref. \cite{makkonen03} for S. As found in earlier work on vicinal
surfaces \cite{stalbov04,makkonen03}, $E_{ad}$ scales roughly
linearly with the number of bonds, $N_{b}$, that the adsorbatemakes
with the substrate, as illustrated in Fig. \ref{figbonds533}. For
Pd(533), the highest $E_{ad}$ is obtained for S or C adsorption at a
4 fold hollow site (the site \# 1). Then, in the order of decreasing
$E_{ad}$ values we have four 3 fold hollow sites (\# 2, 3, 4 and 5),
three 2 fold bridge sites (\# 6, 7, and 8) and two on-top sites (\#
9 and 10). The physics behind this trend can be understood in terms
of the tight binding approximation in which the $p$C-$d$Pd band
width is proportional to the number of nearest neighbors. Broadening
the band leads to depopulation of the anti-bonding $p$C-$d$Pd states
and this way makes the bonds stronger. Some scattering of the
results seen in Fig. \ref{figbonds533} can be attributed to the fact
that the notion of interatomic bond is not well-defined, especially
for such complex geometries as considered in the present work. For
instance, formally carbon adsorbed at the site \#6 has two
neighboring Pd: SC and CC, but, in fact, it also experiences some
weak chemical bonding with two BNN atoms. These extra bonds increase
the adsorption energy and produce a deviation from the linear
dependence. Similarly, C adsorbed at the site \#1, has 5 bonds: four
bonds to the SC and CC Pd atoms which are almost equal, and a fifth
bond to BNN which is longer and causes a downward shift of $E_{ad}$.
Nevertheless, the obtained proportionality of $E_{ad}$ to $N_{b}$
can be used for rough estimation of adsorption energies of C or S on
stepped or kinked Pd surfaces, while deeper insight into the nature
of these phenomena can be gained from analysis of the electronic
structure.

It is interesting to compare the adsorption energy for C on Pd(533)
with those on Pd(211)--two vicinal surfaces with same terrace
geometry but different terrace width. We find from Table
\ref{ead533} here and Table I in Ref. \cite{stalbov04} that the
respective adsorption energies are very similar, indicating that the
step-step separation on Pd(211) itself is large enough so as to not
affect the C binding energy to the Pd atoms.

\subsection{Electronic Structure}

Using the FLAPW method, we have calculated the valence charge
densities and local densities of electronic states for Pd(533) and
Pd(320) with S and C adsorbed on the site labeled \# 1 in Fig.
\ref{fig533320}. To understand the character of chemical bonding
between C, S and metal surface atoms, we have plotted the valence
charge densities along the planes including the most important C-Pd
or S-Pd bonds. Projections of these planes on the (533) surface are
schematically shown in Fig. \ref{figchgdenplanes533}. Note that as a
result of the complex adsorption geometry, centers of some atoms
appear to be slightly out of the planes in the figure. Contour plots
of valence charge densities along these planes are shown in Fig.
\ref{figchden533}. We cut off high densities around the atomic cores
to show in detail the most important low density charge distribution
in interstitial regions. Although in Figs. \ref{figchden533}a and
\ref{figchden533}c the centers of the C and S atoms are not in the
plane of interest here, the charge density "bridges" indicating
covalent bonds are clearly seen between the adsorbates and the SC
and CC atoms. The densities shown in Figs. \ref{figchden533}b and
\ref{figchden533}d reflect the above mentioned differences in the
location of C and S on Pd(533): the C atom is linked to BNN, while
the S atom is located far from it, but close to the TC2 atom.
Further, the intense charge density bridge connecting C and BNN in
Fig. \ref{figchden533}b suggests a strong covalent bonding between
these atoms, while for the same token the S-TC2 bond appears to be
weaker, as seen from Fig. \ref{figchden533}d.

More details about chemical bonding can be obtained from plots
showing the difference $\delta \rho \left( r\right) $ between the
self-consistent charge density of the system and the sum of
densities of free atoms placed at the corresponding sites. Such
plots reflect the charge redistribution caused by chemical bonding.
Figs. \ref{figchdendiff533}a and \ref{figchdendiff533}b show the
$\delta \rho \left( r\right) $ calculated for C/Pd(533) and
S/Pd(533) plotted along the planes (b) and (d) as defined in Fig.
\ref{figchgdenplanes533}. It is seen from the figures that bulk-like
Pd atoms (located comparatively far from the surface) donate some
electronic density to interstitial region to build comparatively
weak Pd -- Pd covalent bonds. Both C and S accept a significant
amount of electronic charge from neighboring Pd atoms, making the C
-- Pd and S -- Pd bonds essentially ionic. However, the distinctive
large electronic density bridge between C and BNN atoms (see Fig.
\ref{figchdendiff533}a) reflects strong C -- Pd covalent bonding. In
contrast, no significant electronic charge density is seen between S
and TC2 atom in Fig. \ref{figchdendiff533}b. Thus, the S -- Pd
bonding in the S/Pd(533) case is mostly ionic with a small covalent
contribution, whereas C and the BNN atom form a mixed ionic-covalent
bond.

Next we turn to the analysis of the local densities of electronic
states (LDOS), which provide additional information about the
character of chemical bonding between adsorbates and metal atoms and
some insights about properties related to catalytic activity of the
surfaces. In Fig. \ref{figdos533} we show LDOS calculated for
S/Pd(533) and C/Pd(533). We find a large splitting of $p$ states of
C and S with two main structures (A and B in the figure) separated
by $(7-8$ eV$)$. Since similar (but less intense) structures are
found in the LDOS of Pd atoms at the same energies, we conclude that
there is a strong hybridization of adsorbate $p$ states with the
local states of surface Pd atoms. The B and A structures thus
represent bonding and anti-bonding states, respectively. As seen
from the figure, the SC, CC and BNN surface atoms are involved in
the hybridization. The B and A structures in the LDOS of carbon are
quite distinct (low density between them) and the anti-bonding
states are empty. These are indications of strong covalent C -- Pd
bonding, in agreement with the obtained valence charge densities. In
contrast, only part of $p$ states of sulfur is involved in the
hybridization (the rest are distributed between B and A structures).
Furthermore, the S anti-bonding states are partially occupied. This
explains why the S -- Pd covalent bonds are weak, as inferred also
from the plots of the valence charge densities in Fig.
\ref{figdos533}.

Similar results are obtained for the adsorbates on Pd(320) (see Fig.
\ref{figdos320}): there is a strong hybridization between the
electronic states of the adsorbate and and neighboring Pd atoms, and
because of larger B -- A splitting and fewer occupation of the
anti-bonding states, carbon forms stronger covalent bonds with Pd
than done by sulfur.

Such significant modification of the electronic structure as we have
documented above for the Pd surfaces upon C and S adsorption should
affect their properties. Since Pd is a widely used catalyst, the
property of interest is its catalytic activity (reactivity).
Recalling the model that links the surface reactivity to the local
densities of electronic states at the Fermi level [$N_a( E_{F})$]
\cite{feibelman84}, we examine the change of this quantity upon C
and S adsorption on the Pd surfaces. As seen from the Figs.
\ref{figdos533} and \ref{figdos320}, the splitting caused by
hybridization reduces dramatically the LDOS around the Fermi-level
for Pd atoms, which have direct covalent bonds with the adsorbate.
In Tables \ref{dos533} and \ref{dos320} we list $N_{Pd}( E_{F})$
calculated for clean Pd(533) and Pd(320), as well as for those
adsorbed with C and S. All atoms of the clean Pd surfaces are found
to have high $N_{Pd}( E_{F})$, which are not much affected by the
presence of the step or kink. This is consistent with the fact that
Pd is an efficient catalyst. For both Pd(533) and Pd(320), the
presence of C and S leads to drastic decrease in $N_{Pd}( E_{F})$
for neighboring Pd atoms, while the next neighbors are only slightly
affected. We thus can expect that both S and C poison surface
reactivity of Pd(533) and Pd(320). Interestingly, for the case of
S/Pd(533) the $N_{Pd}( E_{F})$ are suppressed for SC, TC2 and CC,
which are all exposed to the surface. On the other hand, for
C/Pd(533), three Pd sites (SC, CC and BNN) are strongly affected,
but only two of them (SC and CC) are exposed to the surface. Since
only actual surface atoms are involved in catalytic reactions,
decrease in $N_a( E_{F})$ of BNN should not affect surface
reactivity. Nevertheless, the effect of C on $N_{Pd}( E_{F})$ of
Pd(533) is remarkable. Again, the effect of the S and C adsorbates
on $N_{Pd}( E_{F})$ of metal atoms in Pd(533) appears to be similar
to that obtained earlier \cite{stalbov04} for Pd(211). However, in
the case of Pd(211) this results in suppression of $N_{Pd}( E_{F})$
for all surface atoms, while in Pd(533), which has wider terrace,
some surface atoms continue to retain high values of $N_{Pd}(
E_{F})$.

\section{Conclusions}

In the present work we have studied from first principles the effect
of adsorption of carbon and sulfur on the geometric and electronic
structures of the stepped surfaces Pd(533) and Pd(320). We find the
surface lattice to be perturbed dramatically in response to the
adsorption. Our calculations show that C adsorbed at the bridge site
at the edge of the Pd chain in Pd(320) penetrates the surface to
form a sub-surface structure. The adsorption energies are found to
be site dependent and ranging from 5.31 to 8.58 eV for C and from
2.89 to 5.40 for S. The S -- Pd and C -- Pd bonding have mixed
ionic-covalent character with prevailing covalency for the C -- Pd
bonds and dominating ionicity for the S -- Pd bonds. The strong
hybridization between adsorbate and metal electronic states results
in large splitting of the bands, which causes a dramatic suppression
of the local densities of states at Fermi-level for Pd surface atoms
neighboring the adsorbate. This effect is expected to poison
catalytic activity of these surfaces. We have compared the results
obtained for C and S chemisorption on Pd(533) with those obtained
earlier for Pd(211) \cite{stalbov04}. These two stepped surfaces
have similar structures, but Pd(533) has one atomic chain wider
terrace. The adsorption energies, surface relaxation patterns and
the effects of adsorbates on the local densities of electronic
states and valence charge densities obtained for both surfaces are
found to be quite similar suggesting the local character of the
adsorbate impact on geometric and electronic structures of Pd
surfaces.

{\bf Acknowledgments} This work was supported by the US-DOE under
grant No. DE-FGO3-03ER15445. We thank H. Freund and G. Ruprechter
for helpful discussions.

\pagebreak
%----------------TABLE 1----------
\begin{table}[tbp] \centering%
\caption{Multilayer relaxation for Pd(533).\label{relax533}}
\begin{tabular}{|c|c|c|c|}
\hline
{\bf $\delta $d}$_{i,i+1}$ & {\bf S/Pd(533)} & {\bf C/Pd(533)} & {\bf %
Pd(533)} \\ \hline
{\bf $\delta $d}$_{1,2}$ & -12.3 & +29.7 & -15.8 \\ \hline
{\bf $\delta $d}$_{2,3}$ & -19.7 & +3.7 & -11.9 \\ \hline
{\bf $\delta $d}$_{3,4}$ & +44.7 & -32.0 & -6.6 \\ \hline
{\bf $\delta $d}$_{4,5}$ & -4.2 & +36.0 & +24.3 \\ \hline
{\bf $\delta $d}$_{5,6}$ & -22.5 & +3.4 & -6.8 \\ \hline
{\bf $\delta $d}$_{6,7}$ & +5.9 & -8.4 & -10.1 \\ \hline
{\bf $\delta $d}$_{7,8}$ & +14.3 & +3.6 & +5.8 \\ \hline
{\bf $\delta $d}$_{8,9}$ & -4.1 & +7.2 & +4.1 \\ \hline
{\bf $\delta $d}$_{9,10}$ & -14.4 & -5.1 & -6.7 \\ \hline
{\bf $\delta $d}$_{10,11}$ & +9.4 & -4.2 & -0.2 \\ \hline
\end{tabular}
\end{table}
%-------------TABLE 2--------------
\begin{table}[tbp]
\centering
\caption{Multilayer relaxation for Pd(320).\label{relax320}}
\begin{tabular}{|c|c|c|c|c|}
\hline {\bf $\delta $d}$_{i,i+1}$ & {\bf S1/Pd(320)} & {\bf
C1/Pd(320)} & {\bf C5/Pd(320)} & {\bf Pd(320)} \\ \hline {\bf
$\delta $d}$_{1,2}$ & -24.4 & +26.4 & +38.1 & -15.4 \\ \hline {\bf
$\delta $d}$_{2,3}$ & -5.5 & -55.7 & +1.0 & -18.7 \\ \hline {\bf
$\delta $d}$_{3,4}$ & -8.3 & +45.5 & -0.8 & +1.4 \\ \hline {\bf
$\delta $d}$_{4,5}$ & +34.9 & -8.5 & +4.1 & -10.1 \\ \hline {\bf
$\delta $d}$_{5,6}$ & -16.4 & +23.9 & -5.6 & +21.3 \\ \hline {\bf
$\delta $d}$_{6,7}$ & -3.2 & -9.6 & +16.5 & -7.0 \\ \hline {\bf
$\delta $d}$_{7,8}$ & +12.4 & +9.6 & -3.6 & -1.7 \\ \hline {\bf
$\delta $d}$_{8,9}$ & -14.1 & -4.8 & +7.9 & +2.4 \\ \hline {\bf
$\delta $d}$_{9,10}$ & +11.5 & +0.8 & -4.4 & -1.4 \\ \hline
\end{tabular}
\end{table}
%----------------TABLE 3------------------
\begin{table}[tbp]
\centering
\caption{C and S adsorption energies at various sites on Pd(533) and Pd(320).
\label{ead533}}
\begin{tabular}{|c|c|c|c|}
\hline {\bf Site} & {\bf S/Pd(533)} & {\bf C/Pd(533)} & {\bf
C/Pd(320} \\ \hline {\bf 1} & 5.40 & 8.44 & 8.58 \\ \hline {\bf 2} &
5.31 & 7.52 & 8.29 \\ \hline {\bf 3} & 5.28 & 7.49
& 8.47 \\ \hline {\bf 4} & 5.28 & 7.49 & 7.30 \\
\hline {\bf 5} & 5.26 & 7.43 & 8.49 \\ \hline {\bf 6} & 5.09 & 7.10
& 5.53 \\ \hline {\bf 7} & 4.62 & 6.32 & \\ \hline
{\bf 8} & 4.72 & 5.85 & \\ \hline {\bf 9} & 4.25 & 5.58 & \\
\hline {\bf 10} & 4.13 & 5.31 & \\ \hline
\end{tabular}
\end{table}
%--------------TABLE 4-----------------
\begin{table}[tbp]
\centering
\caption{Local density of states at Fermi level (state/eV*atoms) calculated for Pd
surface atoms in clean Pd(533), S/Pd(533), and C/Pd(533).\label{dos533}}
\begin{tabular}{|c|c|c|c|c|c|}
\hline
& {\bf SC} & {\bf TC1} & {\bf TC2} & {\bf CC} & {\bf BNN} \\
\hline
{\bf clean-Pd(533)} & 2.01 & 1.95 & 2.00 & 1.78 & 1.93 \\ \hline
{\bf S/Pd(533)} & 0.93 & 1.73 & 1.30 & 0.90 & 2.04 \\ \hline
{\bf C/Pd(533)} & 0.37 & 1.78 & 1.98 & 0.46 & 0.65 \\ \hline
\end{tabular}
\end{table}
%----------------TABLE 5--------------
\begin{table}[tbp]
\centering
\caption{Local density of states at Fermi level (states/eV*atoms) calculated for Pd
surface atoms in clean-Pd(320), S/Pd(320), and C/Pd(320).\label{dos320}}
\begin{tabular}{|c|c|c|c|c|}
\hline
& {\bf Pd1} & {\bf Pd2} & {\bf Pd3} & {\bf Pd4} \\ \hline
{\bf clean-Pd(320)} & 1.59 & 2.26 & 1.81 & 1.88 \\ \hline
{\bf S/Pd(320)} & 1.13 & 0.73 & 0.83 & 1.63 \\ \hline
{\bf C/Pd(320)} & 0.44 & 1.27 & 0.24 & 1.30 \\ \hline
\end{tabular}
\end{table}
\end{document}